\documentclass[preprintnumbers,amsmath,amssymb,floatfix,superscriptaddress]{revtex4}
\usepackage{epsfig}
\usepackage{graphicx} 
\usepackage{bm}       
\usepackage{color}
\usepackage{epsfig}
\usepackage{subfigure}
\usepackage{epstopdf}

\def\>{\rangle}
\def\<{\langle}
\def\be{\begin{equation}}
\def\ee{\end{equation}}
\def\bee{\begin{equation*}}
\def\eee{\end{equation*}}

\begin{document}

\title{Scale-estimation of quantum coherent energy transport in multiple-minima systems}

\author{Tristan Farrow}
\email{t.farrow1@physics.ox.ac.uk}
\affiliation{Atomic and Laser Physics, Clarendon Laboratory, University of Oxford, OX1 3PU, UK.}
\affiliation{Centre for Quantum Technologies, National University of Singapore, Singapore 117543.}
\affiliation{Oxford Martin School, Old Indian Institute, University of Oxford, OX1 3BD, UK.}
\author{Vlatko Vedral}
\affiliation{Atomic and Laser Physics, Clarendon Laboratory, University of Oxford, OX1 3PU, UK.}
\affiliation{Centre for Quantum Technologies, National University of Singapore, Singapore 117543.}
\affiliation{Oxford Martin School, Old Indian Institute, University of Oxford, OX1 3BD, UK.}

\date{\today}
\maketitle

\textbf{A generic and intuitive model for coherent energy transport in multiple minima systems coupled to a quantum mechanical bath is shown. Using a simple spin-boson system, we illustrate how a generic donor-acceptor system can be brought into resonance using a narrow band of vibrational modes, such that the transfer efficiency of an electron-hole pair (exciton) is made arbitrarily high. Coherent transport phenomena in nature are of renewed interest since the discovery that a photon captured by the light-harvesting complex (LHC) in photosynthetic organisms can be conveyed to a chemical reaction centre with near-perfect efficiency. Classical explanations of the transfer use stochastic diffusion to model the hopping motion of a photo-excited exciton. This accounts inadequately for the speed and efficiency of the energy transfer measured in a series of recent landmark experiments. Taking a quantum mechanical perspective can help capture the salient features of the efficient part of that transfer. To show the versatility of the model, we extend it to a multiple minima system comprising seven-sites, reminiscent of the widely studied Fenna-Matthews-Olson (FMO) light-harvesting complex. We show that an idealised transport model for multiple minima coupled to a narrow-band phonon can transport energy with arbitrarily high efficiency.
}

\section{Introduction}

\medskip

We use the principle of minimum design to develop a general mechanism whereby energy can be transported with near-perfect efficiency through a generic system with multiple energy minima. Spectroscopists have been measuring wavelike electronic quantum coherences in photosynthetic light-harvesting protein complexes~\cite{fleming04, fleming05, fleming07sci, col10, engel10, engel11, engel13}, including at the single-molecule level~\cite{hulst13}, since 2004. The molecules capture photons from sunlight and transport their energy via photoexcited excitons to a chemical reaction center with near-perfect efficiency, unmatched by artificial systems, by computing the fastest path to the reaction centre. These observations add weight to the prediction articulated by the founders of quantum physics, Werner Heisenberg and Erwin Schr\"odinger, that a new picture of biology would emerge when quantum theory would be applied to its study~\cite{schrod, heis, ball, vlatko09}.

\medskip
Considering that primitive photosynthetic cells (Stromatolites) appeared over three billion years before any other lifeform, we can speculate that nature assembled a photosynthetic mechanism with minimal resources and evolved it to near-perfection under the action of natural selection. Land-based plant life appeared much later, around 400 million years ago, and is a  recent development in the evolutionary history of photosynthesis. Seen in this light, the high efficiency of energy-transfer in light-harvesting is perhaps less surprising.

\medskip
From the perspective of quantum physics, biomolecules are large and hot objects, so at first sight, it seems surprising that quantum mechanics might play a useful role. It is beyond the scope of this paper to speculate on the functional value of quantum processes in biology, however, we can look to the timescales of biological processes for insights to inform our analysis. Typically, we have four distinct timescales, each separated by around three orders of magnitude. The fastest biological processes occur on the timescale of optical excitations, typically femtoseconds ($10^{-15}$s). The next timescale covers the coherent hopping of excitons, of the order of picoseconds ($10^{-12}$s). This is followed by chemical reactions that deliver the excitonic energy to say the ATP cycle, which then occurs on timescales of microseconds. Finally, there are the macroscopic timescales from milliseconds to seconds, in processes like neurological response.

\medskip
It follows from a heuristic argument that quantum coherence can realistically only be present within the first two timescales~\cite{leg}. Consider a typical molecule of at least one hundred thousand atoms, which absorbs a photon that projects the molecule into a conformal quantum superposition of different spatially arranged states. Such conformal changes are ubiquitous in living molecules.  How long can this superposition survive? Analysis suggests that the ratio of decoherence time to dissipation time is $\hbar^2 / (mx^2 kT)$, where \emph{m} is the mass of the molecule, \emph{x}, is the coherence length, a few nanometers and comparable to the molecular size, and \emph{T} is the temperature, $\sim 300~K$. The dissipation, then, is typically of the order of seconds to milliseconds, which in turn leads to decoherence times of nano- to pico-seconds. A more detailed and rigorous analysis leads to the same conclusion~\cite{leg}. So only processes on similar or shorter timescales can survive the environmental interaction and persist longer than what might be expected of processes that decohere before they affect the dynamics. These two timescales, in the femto- and picosecond domains, are exactly those relevant to the process responsible for the efficient transfer of energy in the LHC.

\begin{figure}[htbp]
\centering
\includegraphics[angle=0, width=8.6cm]{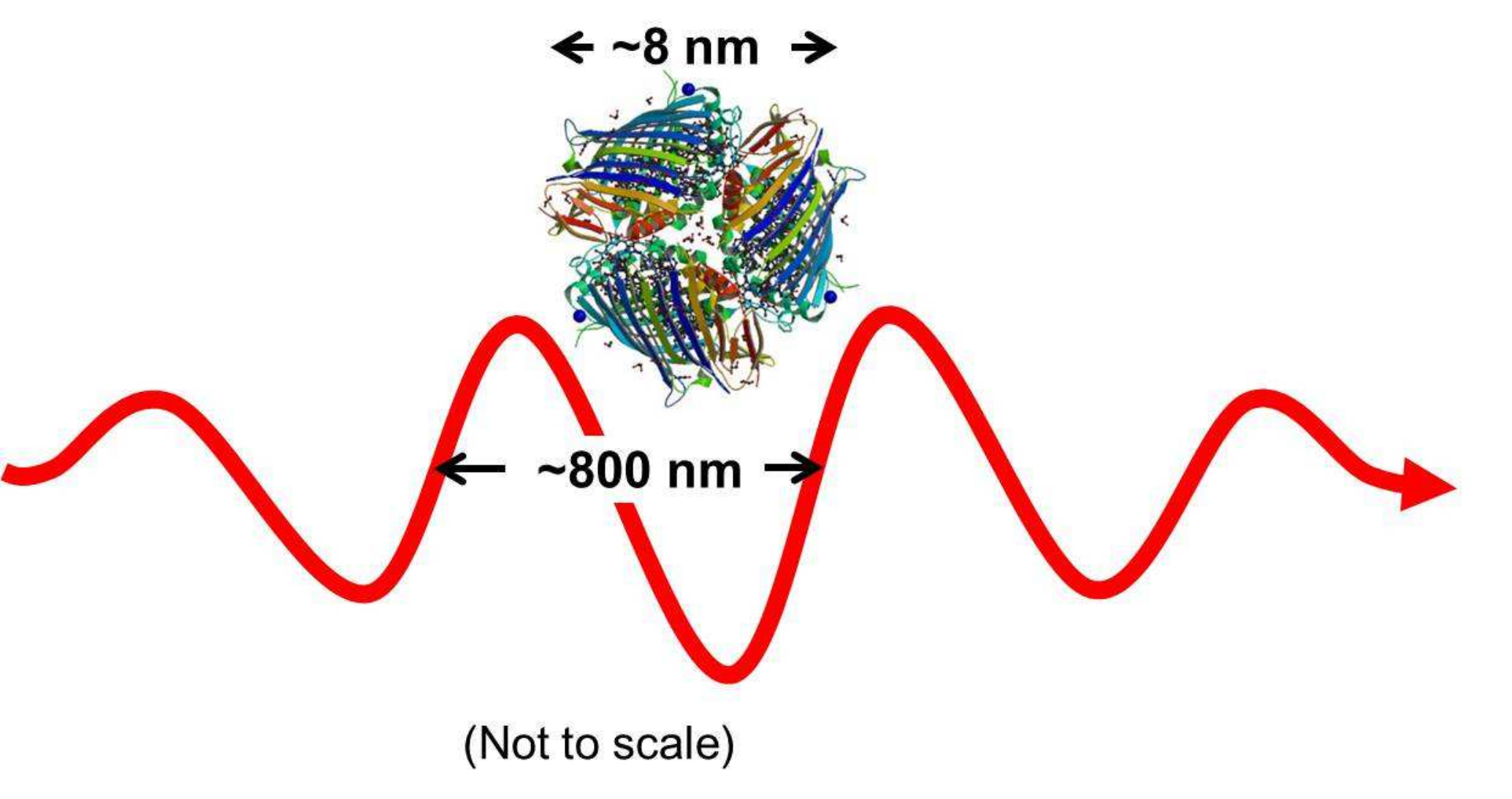}
\caption{Light-harvesting pigment protein of the green sulphur photosynthetic bacterium $Prosthecochloris$ $aestuarii$~\cite{blanken} that lives in marine environments. The wavelength of the incoming photon field, $\sim 800~nm$, is two orders of magnitude larger than the dimensions of the molecule.}
\label{fmo}
\end{figure}

\medskip
A widely used tool for calculating energy transfer rates between molecules is given by F\"orster theory~\cite{forster}. This yields a rate for the energy transport from the overlap between the absorption and emission spectra of donor and acceptor states. In the semi-classical picture, as presented by F\"orster theory, energy is transferred by the incoherent hopping of photoexcited excitons down an energy ladder towards a chemical reaction center.

\medskip
The nano-meter dimensions and dense packing of the LHC pigment sites, where intermolecular separation is of the order of the size of the molecules, allow us to postulate that the protein environment could indeed host strong correlations associated with quantum phenomena. The picture that emerges then is quantum mechanical in that excitons display a wavelike character and exist in a quantum superposition of states. This allows an exciton to enter simultaneaously different energy pathways before finding the fastest route to the reaction center, so that no energy is lost. That is the picture suggested by the recent experiments on LHCs~\cite{fleming04, fleming05, fleming07sci, col10, engel10, engel11, hulst13, engel13}. This is at odds with F\"orster theory, which cannot readily account for coherence between the donor and acceptor states. However, without quantum effects, the rapid energy transfer cannot be satisfactorily explained either.

\medskip
For example, the FMO molecule of bacteriochlorophyll has been the center of attention in recent years. This consists of seven trapping sites each able to hold an exciton~\cite{scholes06}. When a photon is absorbed by the FMO protein, a single electron-hole pair is photoexcited in the complex. F\"orster theory, where the exciton hops from site to site until it finds the reaction center where it deposits its energy, fails to account for the hundreds of femtosecond timescale~\cite{fleming07nat, vlatko09} measured for the process. Experiments~\cite{fleming04, fleming05, fleming07sci, col10, engel10, engel11, hulst13, engel13} and numerical simulations~\cite{ishi} suggest a coherent superposition of a single exciton state over the seven sites of a photosynthetic pigment. From a quantum physical perspective, the absorption of a photon, whose wavelength is 100 times larger than the length-scale of the photosynthetic aparatus (Fig.~\ref{fmo}), projects the excited state into a quantum superposition of states spread over seven sites of different energy and with a different probability amplitude of occupation.

\medskip
Just before we consider the details of our generic model, let us briefly put it into context. Existing proposals~\cite{castro08, plenio1, lloyd09} resort to stochastic theory and introduce Markovian statistics. We adopt a nonmarkovian analysis that avoids this complexity to reveal the underlying transfer mechanism in an intuitive manner. A nonmarkovian approach was also explored in~\cite{ishi}, which resorts to additional complex features redundant in our model. Furthermore, the efficiency predicted in theory~\cite{lloyd09} is lower than experimentally measured values and appears to diverge at higher temperatures from what might be expected from a system designed to shield quantum coherence. Coherent random walk proposals~\cite{castro08, plenio1, lloyd09} approach the problem of energy transfer from the viewpoint that the probability of transfer depends on the exciton being ejected from energy minima where it comes unstuck, through coupling to a thermal bath. The trapped excitonic states, or dark states, then rely on a coherent random walk to reach the reaction center. This contrasts with the minimalistic mechanism we present here, which dispenses with noise.

\medskip
The nonmarkovian property of the system and phonon environment dynamics can formally be phrased as follows. Any Markovian dynamics have the property that the distinguishability between input states cannot increase with time, and in fact, it usually decreases. If, then, we record the opposite trend in the evolution of our system, namely that the distinguishability of states increases, the immediate conclusion is that the system-to-environment coupling is nonmarkovian. The increase in distinguishability is therefore sufficient to witness nonmarkovianity. Our model employs two qubits coupled to a narrow-band phonon mode. It is nonmarkovian in that the ability to distinguish two qubit states can increase with time. This means that information flows coherently from the environment back into the system, contrary to Markovian processes. However, in our semiclassical approximation, where the phonon mode is treated as a static classical field, the ensuing evolution of the two qubits is unitary. Another way of saying this is that our vibrational modes can in fact be thought of as embedded within the system, and instead of introducing noise, they promote coherence. Existing proposals assume either markovian or nonmarkovian environments. We believe instead that neither of these two statistical properties of the environment are necessary to explain the energy transfer efficiency.

\medskip
A clearer understanding of the energy transfer process in complex molecules is obscured by ongoing debates surrounding the exact nature of and coupling to thermal noise in the environment. Much work has been done on modelling this in biomolecules such as FMO~\cite{shim12, reich12} through atomistic simulations, but at present there is insufficient experimental data to conclusively characterise or identify the phononic or vibronic profiles believed to aid transport. In contrast, the authors of~\cite{fleming07nat} conclude that a non-standard phononic environment might be responsible for preserving the coherence seen in their experiments. A system that seeks to preserve coherence would, it seems reasonable, seeks to optimise or exclude coupling to a thermal bath. In the ideal and simplest case, a single-mode would be sufficient to bring into resonance energy-detuned sites.

\medskip
What we do know, however, is that regardless of the exact profile of the environment, noise-assisted transport requires one or several phonon modes of appropriate frequency to bridge the energy gaps in any given system. Such a mechanism is ubiquitous in nature, as in generic double-minimum systems interacting with a quantum mechanical heat bath, and in transfer kinetics in condensed hydrogen-bonded systems~\cite{meyer90, heuer91}. We underline that a realistic phononic environment is likely to present a more complicated mode structure, but our purpose here is to distil only the salient features of the transport problem to offer a fresh and intuitive understanding of how phonon-aided transport can enhance energy transfer.

\medskip
We take as our starting point the simplest case of a donor-acceptor system coherently coupled to a phonon mode with a single optimal frequency in the limit where the coupling is weaker compared to the energy gap between the sites ($J << \Delta$). In the absence of further experimental evidence, this approach offers an intuitive insight that complements other approaches that build a case from the 'ground-up' through atomistic simulation~\cite{shim12}. The single-mode approach in particular allows us to neglect dephasing and to adopt a purely quantum mechanical treatement that illustrates the salient features of the transport mechanism.

\medskip
We offer a heuristic argument for our approximation of the coupling rate between individual donor and acceptor sites relative to their detuning and phonon energy (in the case of FMO). The mechanism for our transport model itself stands independently of these approximations and can be extended to other multiple minima systems.

\section{Two-level toy system}

\begin{figure}[htbp]
\centering
\includegraphics[angle=0, width=11cm]{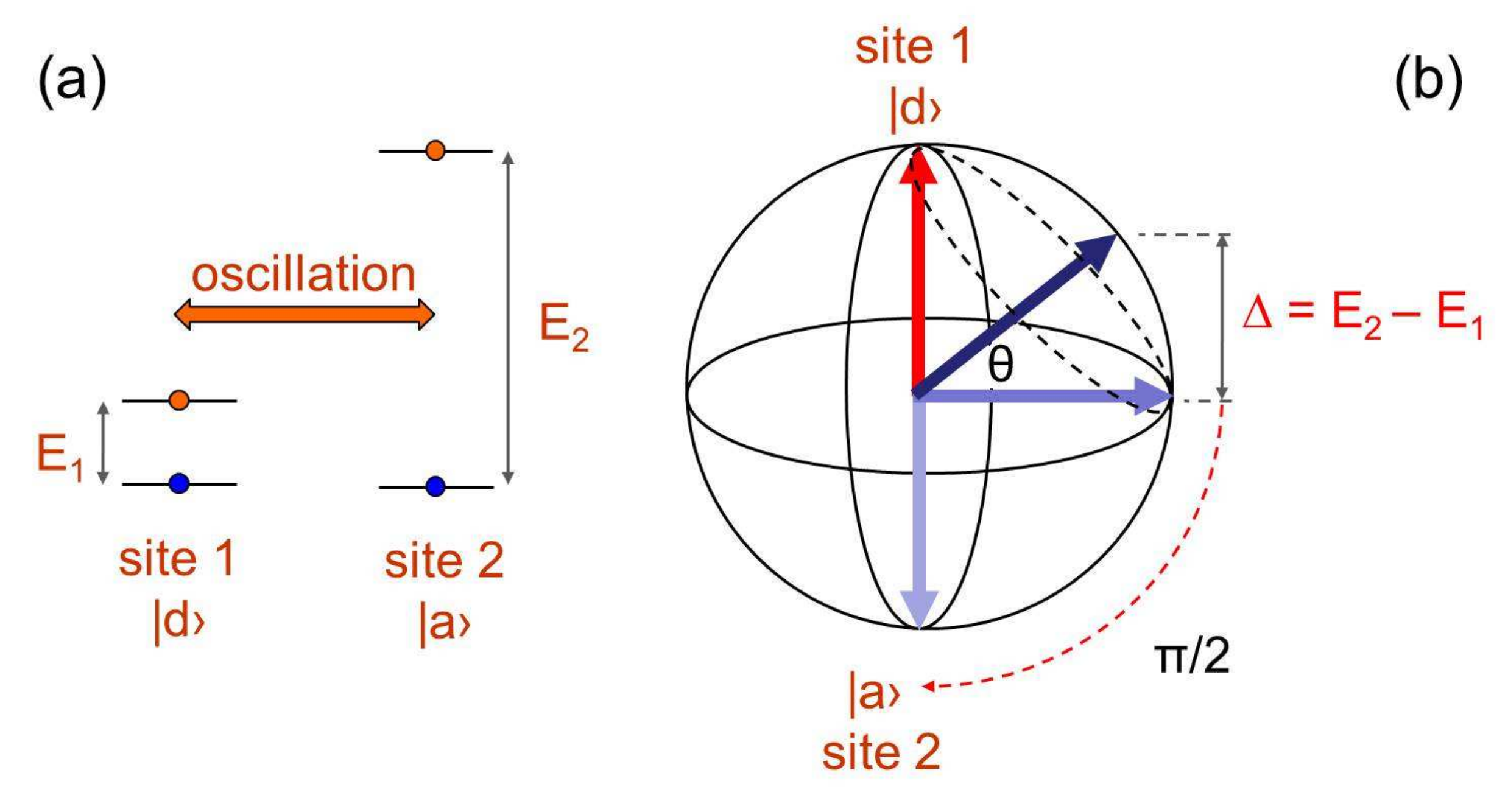}
\caption{(a) Vibrationally-assisted transition between two sites, 1 and 2, representing the exciton~\cite{scholes06} donor and acceptor states, $|d\>$ and $|a\>$, respectively.  with an energy mismatch. (b) Bloch sphere representation of the transition, illustrating the detuning from resonance between the two states in the two sites. On transition between the two sites, the Bloch vector flips between the two sites which are detuned by an energy deficiency $\Delta$. The parameters are: $\Delta$, the coherence restitution term, where $\Delta = E_2 - E_1$, and $E_2 - E_1$ is the energy difference between the excited states of sites $1$ and $2$. On resonance, $\Delta = g |\alpha|$, which is analogous to Rabi oscillations in atoms.}
\label{bloch}
\end{figure}

The first mechanism we consider is a donor-acceptor system with F\"orster dipole-dipole exchange interaction. The problem we address is the efficiency of exciton transfer from a donor state, $|d\>$, to an acceptor, $|a\>$, given that the two are generally widely detuned in energy, as in Fig.~\ref{bloch}(a), illustrating two energy-detuned sites 1 and 2. The detuning $\Delta$ is typically of the order of optical frequencies, while the hopping interaction strength, $J$, due to F\"orster coupling, is some two orders of magnitude smaller. In the absence of another mechanism, the excitation would have little chance transferring efficiently from the donor to the acceptor. A straightforward way of illustrating this is to focus only on the relevant subspace, by reducing the system to a two-level toy system. One level designates the donor energy, and the other the acceptor energy, the two separated by the detuning $\Delta$ and driven at the rate $J$. This is the same as the Rabi model of a two-level system driven by an external field. The Bloch sphere representation in figure~\ref{bloch}(b) illustrates the $\theta= \pi/2$ rotation of the Bloch state-vector (indicated by the arrows). This rotation, hence the evolution of the system from the donor site $|d \>$ to the acceptor site $|a\>$, is characterised by the well- known probability of transfer:
\medskip
\be
p= \frac{J^2}{J^2+\Delta^2}
\label{proba}
\ee

\medskip
Given that $J$ is roughly ten to a hundred times smaller than $\Delta$, the probability is as small as $10^{-2} - 10^{-4}$. When we consider the near unit efficiencies of energy transfer light-harvesting, this mechanism is insufficient. In practice, the donor-acceptor system is part of a complex molecule that has a multitude of vibrational degrees of freedom, in addition to the electronic component. There are usually acoustic vibrations (of order megahertz) and optical vibrations, around three orders of magnitude higher. Given that the transfer can take place at the rate of inverse $J$ (on picosecond time scales), acoustic vibrations will be irrelevant, while optical ones play a useful role. In our system, the initially detuned donor and acceptor are shifted into resonance by a narrow band of phonons that are coherently excited.

\medskip
Typically, phonons couple to excitons in an off-resonant and dispersive fashion. Normally, we think that this naturally dephases the excitons. However, this is only really true if, generally speaking, the phonons have a broad spectrum and are subject to weak coupling to excitons. This leads to the regime given by the Born-Markov approximation~\cite{gard}. If we have a strongly coupled narrow band of phonons instead, then the vibrational degrees of freedom act coherently and can in fact aid the transfer. This is the key mechanism we are proposing.

\medskip
To see how phonons can enhance the exciton transfer, let us introduce a system of vibrations. We start with the full Hamiltonian, which includes the donor and acceptor, the phonons as well as electron-phonon coupling:

\medskip
\begin{equation}
H = \frac{E_1}{2} \sigma_d^z + \frac{E_2}{2} \sigma_a^z + J (\sigma_d^x\sigma_a^x+\sigma_d^y\sigma_a^y) + \sum_n \hbar \omega_n a^{\dagger}_n a_n + \sum_i (g^1_i \sigma^z_1 +g^2_i \sigma^z_2)(a_i+a_i^{\dagger})
\end{equation}

\medskip
The energy difference between donor and acceptor is $E_2-E_1 = \Delta$, the exchange coupling is $J$, the third term is the phonon energy and the last is the electron-phonon coupling. Now we make four assumptions, which define our system:

\begin{enumerate}
\item	The phonon density is narrow, so that we can approximate relevant vibrations by a single phonon mode.
\item	The donor and acceptor are exactly out of phase in their coupling to vibrations.
\item	The vibrational mode can be treated classically. This means that it can be treated as a coherent state (or near-coherent) of amplitude $\alpha$. Thermal states are mixtures of coherent states, so they also satisfy this condition.
\item	The frequency of phonons is much smaller than J (quantified below).
\end{enumerate}

The Hamiltonian representing the donor-acceptor qubit coupled to a single vibrational mode (assumption 1) is given by:
\begin{equation}
H = \frac{\Delta}{2} (|d\rangle\langle d| - |a\rangle\langle a|)  + J (|d\rangle\langle a| - |d\rangle\langle a|) + \hbar \omega a^\dagger a + (g_1\sigma^z_1 +g_2 \sigma^z_2)(a+a^\dag)
\label{ham}
\end{equation}

Here $\omega$ is the phonon frequency, and $g_d$ and $g_a$ are the exciton-phonon coupling strengths for the donor and acceptor respectively. Assumption 2 states that the coupling is of opposite phase, which means that, $g_d=-g_a=g$. This is of central importance to our mechanism since this means the energy shift acts in such a way as to bring the donor and acceptor into resonance, provided that the product of the phonon coupling strength and its amplitude of oscillation is equal to the detuning. Implementing assumption 3 means rewriting the last term as $g \alpha (\sigma^z_d - \sigma^z_a)$, where $\alpha$ is the real part of the amplitude of the phononic coherent state. Because we assume that phonons evolve slowly compared to relevant time scales (assumption 4), the amplitude is practically time-independent. The resulting Hamiltonian now represents effectively a DC Stark shift in excitons, induced by phonons:
\medskip
\be
H = \begin{matrix}
\phantom{....} \<d| \qquad \qquad \<a|\\
\begin{matrix}
|d\> \\ |a\>
\end{matrix}
\begin{bmatrix}
\Delta-g |\alpha| & 2J \\
2J & -\Delta+g |\alpha| \\
\end{bmatrix}
\end{matrix}
\ee

\medskip
If $\Delta = + g\alpha$, then the donor-acceptor system is DC Stark-shifted into full resonance. This could be the modulation postulated in~\cite{fleming07nat}, namely that the protein could be modulating the electron-phonon coupling. Now that the transfer due to the ensuing Rabi oscillations becomes resonant, it follows that the probability of transfer reaches unity. In practise, that probability will of course be imperfect. So let us consider how accurately each of the assumptions must hold to maintain a high fidelity of excitation transfer.

\medskip
There are four main types of errors in our system, all stemming from the approximations introduced in the vibrational degrees of freedom. One is that the phonons will most certainly not be single-mode, but will instead have some spread, which we label $\delta \omega$. Each of these modes will couple with a different strength to excitons, say $g_i$, where $i=1,2,...,7$ is the FMO  site index. The second error could come from the fact that the donor and acceptor are not exactly out of phase in the phonon mode. The third type of error comes from the spread in phonon number in each mode. Our mechanism works ideally with phonon states that are near number-squeezed, in the sense that the dispersion in phonon numbers does not exceed the root of the mean number of phonons. This is why we use coherent states in our calculations, although phase-coherence is by no means prerequisite and the corresponding number-state mixture would also suffice to achieve the same evolution. This is why thermal states can also achieve a high efficiency. Finally, the fourth type of error derives from the assumption that the phonon modes are static, i.e. that the oscillation frequency is much smaller than $J$. In effect, this means taking $\cos \omega t$ to be equal to unity. To second-order approximation, this error equals $\omega^2 t^2$, which contributes $g \alpha \omega^2/J^2$ to the detuning between excitons. We can illustrate the robustness of the system with the following general argument.

\medskip
It can readily be seen from equation~(\ref{proba}), that for a transfer efficiency of $75\%$ (i.e. an efficiency drop of $25\%$), the error on $\omega$ including all possible sources needs to be as high as $50\%$ of the oscillation frequency, $\omega$, since $J \sim 10^{13}$Hz and $p \sim 1 - (\delta \omega / J)^2$, where $\delta \omega$ is the error on $\omega$. It follows that the mechanism provides ample overhead on errors, since it tolerates inaccuracies of up to $50\%$ in $J$, while maintaining the on-resonant conditions that lead to the very high efficiencies of energy transfer. Importantly, we note that the highest classical limit for quantum state transfer is $\frac{2}{3}$~\cite{Massar}, so even with a $50 \%$ error, our mechanism shows that quantum coherence still contributes amply to the efficiency of energy transfer.

\medskip
The intuitive formulation of our system makes it accessible to experimental tests. For example, the assumption that the FMO protein maintains a narrow phononic band can be verified in tests aiming to reveal a high degree of temperature insensitivity of the phononic mode. This stands to reason since our mechanism is based on processes that occur at frequencies in the optical and near optical regime, whose energy is well above $kT$ at room temperature and below. Although the latest experimental results do not seek to ascertain the profile of the phonon modes as a function of temperature, they do confirm that coherence persist even at ambient temperature~\cite{col10}. This supports the speculation view that the FMO protein could have evolved a specialised mechanism to modulate the electron-phonon interaction~\cite{fleming07nat}.

\section{A seven-site system}

\begin{figure}[htbp]
\centering
\includegraphics[angle=0, width=8.2cm]{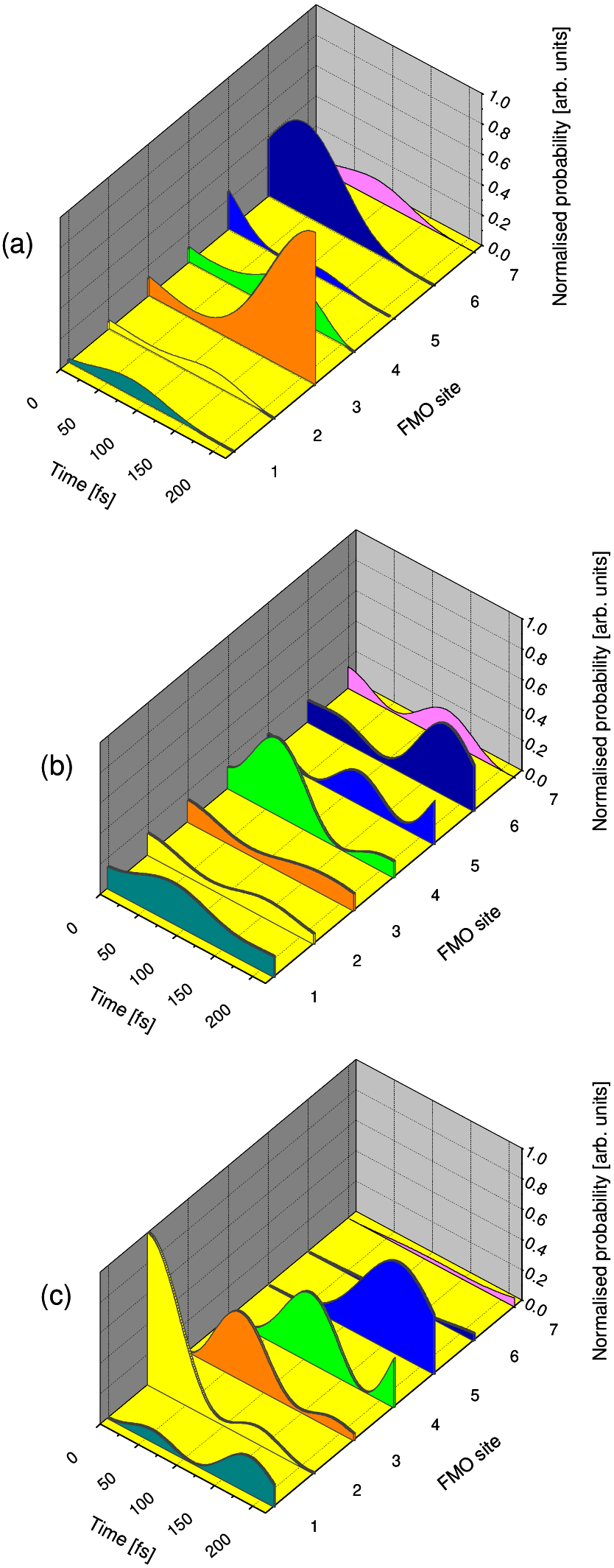}
\caption{Simulated time-evolution of the probability distribution of a quantum superposition of an exciton's wavefunction across sites 1 to 7 of the FMO complex in $Prosthecochloris$ $aestuarii$. (a) shows an ideal quantum superposition leading to excitonic confinement in site 3 (designated reaction centre) after $\sim 220$~fs. (b) shows the evolution for an initial state in an equal superposition of the exciton in all seven sites, and (c), for an arbitrarily chosen initial pure state in site 2. The curves were obtained by plotting the relevant diagonal elements extracted from the evolution of the full state, including all the off-diagonal elements.}
\label{evolution}
\end{figure}

\medskip
We illustrate the generality of our mechanism by applying it to the seven-site system in the light-harvesting complex of $Prosthecochloris$ $aestuarii$, a green sulphur photosynthetic bacterium (Figure~\ref{fmo}) living in marine environments. The Hamiltonians of many FMO complexes and of $Pr.$ $aestuarii$ in particular have been extensively studied and their energies catalogued. The Hamiltonian we use here is represented by a seven-dimensional matrix~\cite{adolphs06, plenio1}.

\medskip
Although the exciton energies, $E_i$, at sites $i=1,2,...,7$, and the hopping rates $J_{ij}$ between sites $i$ and $j$ are well characterised, we use a range of relevant phonon frequencies and their exciton coupling strengths. As before, we assume a narrow-band of vibrational modes to bring the seven sites into resonance with each other. Even though our mechanism remains minimalistic, it is difficult to treat analytically.

\medskip
The time-evolution of the probability amplitudes of a quantum superposition of an exciton in seven sites in the $Pr.$ $aestuarii$ light harvesting protein was simulated numerically. Figure~\ref{evolution}(a) shows the evolution resulting in the capture of the exciton by site 3 within $\sim~200$~fs with 99.99\% probability. Once the exciton is confined to site 3, its energy is deposited in the reaction center. We find, however, that when the excitonic wave-function is superposed equally over all seven sites, figure~\ref{evolution}(b), the probability of confining the exciton in site 3 fails to reach unity, even at multiple periods of the evolution on the timescales of relevant biological processes. A similar outcome is shown in figure~\ref{evolution}(c), where the system is initialized with an exciton localised in a single site, here chosen arbitrarily at site 2, since our simulations showed that this outcome was site-independent. 

\medskip
Our results further indicate that for the $Pr.$ $aestuarii$ FMO complex, the excitonic wavefunction favours site 6, which has a higher occupation probability, at $\sim~20$\%. This is to be expected considering that the exciton energies are different and therefore couple with different strengths to sunlight. Current experimental data cannot for now confirm with greater certainty what the initial state in the FMO complex is, but our analysis suggests that it would be surprising to find all sites excited with the same quantum mechanical amplitude. Although the energetic distribution of seven sites appears to be random and asymmetric, we can reasonably speculate that the FMO protein has evolved in such a way as to capture photons with an energy likely to generate an exciton in an optimal superposition, that is, one leading to efficient energy transfer.

\section{Discussion and conclusion}

\medskip
The question as to how the exciton is trapped in the reaction center remains open. This is a separate but related problem, since without trapping, the excitation will forever oscillate between sites under the action of the unitary evolution. Our mechanism could, with minor modification, tackle the problem of reversibility. When the exciton is captured by the site at the reaction center, a mechanism inspired by the physics of quantum dots can be used, whereby an electron flopping between two potential traps can be confined in one when a rapidly oscillating field of large amplitude is applied at the potential boundary ~\cite{naz}. Whether this mechanism, which is analogous to conventional dephasing, also occurs in FMO complexes is open to further investigation, but it is forseeable that the requirement on the timing of dephasing need not be stringent.

\medskip
The appeal of the mechanism presented here is that it offers an intuitive system that predicts the effects observed in experiments on LHC complexes, and achieves this without resorting to Markovian dynamics or quantum entanglement~\cite{whaley09, vlatkoent, plenio2}. It also seems plausible that such a mechanism could be engineered into biomimetic artificial media such as low-dimensional heterostructures with applications in light-harvesting and efficient energy transport.

\bigskip
\noindent{\bf Acknowledgments}
\medskip

The authors thank Dr Wonmin Son and Dr Kavan Modi for insightful comments and gratefully acknowledge financial support from the Oxford Martin School Programme on Bio-Inspired Quantum Technologies, the Singapore Ministry of Education and National Research Foundation, the UK EPSRC, the Royal Society and the Wolfson Trust. V.V. is Fellow of Wolfson College, Oxford, and T.F. is James Martin Fellow of the Oxford Martin School.


\end{document}